# Brewster quasi bound states in the continuum in all-dielectric metasurfaces from single magnetic-dipole resonance meta-atoms

Diego R. Abujetas[1], Ángela Barreda[2], Fernando Moreno[2], Juan J. Sáenz[3], Amelie Litman[4], Jean-Michel Geffrin[4*], José A. Sánchez-Gil[1*]

**Bound states in the continuum (BICs) are ubiquitous in many areas of physics, attracting especial interest for their ability to confine waves with infinite lifetimes. Metasurfaces provide a suitable platform to realize them in photonics; such BICs are remarkably robust, being however complex to tune in frequency-wavevector space. Here we propose a scheme to engineer BICs and quasi-BICs with single magnetic-dipole resonance meta-atoms. Upon changing the orientation of the magnetic-dipole resonances, we show that the resulting quasi-BICs, emerging from the symmetry-protected BIC at normal incidence, become transparent for plane-wave illumination exactly at the magnetic-dipole angle, due to a Brewster-like effect. While yielding infinite Q-factors at normal incidence (canonical BIC), these are termed Brewster quasi-BICs since a transmission channel is always allowed that slightly widens resonances at oblique incidences. This is demonstrated experimentally through reflectance measurements in the microwave regime with high-refractive-index mm-disk metasurfaces. Such Brewster-inspired configuration is a plausible scenario to achieve quasi-BICs throughout the electromagnetic spectrum inaccessible through plane-wave illumination at given angles, which could be extrapolated to other kind of waves.**

Bound states in the continuum (BICs) are ubiquitous in Physics as a general ondulatory phenomenon[1]. All such states exhibit the common fascinating property that they remain localized despite coexisting with the continuum of radiating channels. First proposed in quantum mechanics, they have been observed in a variety of waves: electromagnetic, acoustic, elastic, water, etc. In the case of electromagnetic waves, BICs are especially relevant for their ability to confine light with infinite lifetimes in geometries for which all radiation channels vanish as a result of destructive interference, while outgoing waves are allowed to propagate in the surrounding medium. Photonic BICs can be implemented in a variety of systems: e.g. photonic crystals[2–10], 1D or 2D arrays of scatterers[11–19], high-Q resonators[20,21], and anisotropic 1D and 2D waveguides[22,23]. Among the various mechanisms that lead to BICs[1], symmetry protection or accidental degeneracy make them especially robust[1]. Robustness in symmetry protected BICs has been connected to their topological nature through the conservation of the topological charge of the associated polarization vortex in the far-field radiation[4,9,12]; whereas robustness in accidental, interferometric BICs stems from the fact that perturbations just shift the BIC frequency without destroying them[14,24]. Nevertheless, there is a lack for means to experimentally achieve electromagnetic BICs with frequency-wavevector space tunability in a simple manner; thus hindering the observation of photonic BICs[7,9].

[1]Instituto de Estructura de la Materia (IEM-CSIC), Consejo Superior de Investigaciones Cientificas, Serrano 121, 28006 Madrid, Spain. [2]Department of Applied Physics, University of Cantabria, Santander, Cantabria 39005, Spain. [3]Donostia International Physics Center DIPC, Paseo Manuel de Lardizabal 4, 20018, Donostia, San Sebastián, Spain. [4]Aix Marseille Université, CNRS, Centrale Marseille, Institut Fresnel, Marseille, France. *Jean-Michel.Geffrin@fresnel.fr; j.sanchez@csic.es

Here we propose to exploit a Brewster-like mechanism to make electromagnetic quasi-BICs inaccessible through plane wave illumination at tunable angles. A planar array is used to impose a single outgoing channel through periodicity, namely, a metasurface. Metasurfaces consisting of plasmonic or dielectric meta-atoms allow for the control of the electric and magnetic field of light which leads to a wealth of phenomenology, thus being promising configurations for planar optical devices[13,25–27]. Recall also that metasurfaces are related to frequency-selective surfaces in the low-frequency regime[28–34], wherein however BICs have never been reported. Lately, the potential of metasurfaces to support a variety of BICs has been explored. BICs at normal incidence (k=0) have been reported[13,35,36]: either due to Friedrich-Wintgen mechanism enforced by asymmetric dimer meta-atoms[35,36]; or due to symmetry protection exploiting meta-atoms (nanodisks) with a complex dipolar response. In the latter, lasing is achieved by modifying one of the lattice constants to allow for an off-normal diffraction channel. Otherwise, it should be mentioned that complex multipolar have been proposed very recently to enforce off-normal cancellation for accidental BICs[17].

In this context, the crucial idea is to combine single-channel metasurfaces with a Brewster-like effect, enforced by using meta-atoms that hold a single non-degenerate dipolar resonance, in turn angularly-locked through e.g. meta-atom orientation. Note that this Brewster effect differs from that obtained through the negative interference between different overlapping dipolar radiation patterns from each meta-atom[37]. The proposed non-degenerate resonances generate a lattice resonant mode that, despite being slightly leaky throughout the k-space, becomes inaccessible for plane wave illumination exactly at the angle at which the (incident or reflected) wavevector and the meta-atom dipole moment are parallel. To this end, we propose high-refractive-index (HRI) subwavelength structures[20,38–44] to play the role of single-resonant meta-atoms exhibiting strong magnetic-dipole resonances. HRI structures have attracted widespread interest as an alternative to Plasmonics[27,43] due to the nearly lossless character and intrinsic nonlinearities of semiconductors (used in the visible as HRI materials). In this work, we demonstrate theoretically and verify experimentally that the above-mentioned mechanism yields such electromagnetic symmetry-protected BICs and Brewster quasi-BICs in the microwave regime; and thanks to the scale invariance of Maxwell equations, directly translatable to the optical regime[45].

**Brewster quasi-BIC concept**

First, we discuss through a simple model the underlying physics. Let us consider a planar periodic array of resonant electric or magnetic dipoles in a frequency regime such that no diffraction orders are allowed (Fig. 1). Such dipoles are characterized by a (electric or magnetic) polarizability tensor $\boldsymbol{\alpha}_0(\omega,\mathbf{k}) \equiv \alpha_0(\omega, \varphi)\mathbf{k}_0$ that is essentially oriented along the dipole moment ($\varphi$ being the dipole moment orientation, see inset in Fig. 1a) and exhibits a resonance at a given frequency $\omega_0$. The expression of the specular reflectivity $R(\omega,\theta)$ for a given angle of incidence $\theta$ can be written as[46]:

$$R(\omega,\theta) \propto |(\cos^2\varphi \sin^2\theta - \cos^2\theta \sin^2\varphi)\alpha|^2, \qquad (1)$$

with $\alpha(\omega,\theta)$ being the dressed polarizability given by:

$$\alpha(\omega,\theta) \propto \left(\frac{1}{\alpha_0} - \widetilde{G}\right)^{-1} \approx \alpha_0/(1 - \alpha_0 \widetilde{G}), \qquad (2)$$

where $\widetilde{G}(\omega, \varphi)$ is the corresponding lattice Green function component projected along the dipole moment angle $\varphi$. If we express the bare polarizability as $\alpha_0 \sim g(\varphi) * \omega_i / (\omega - \omega_0 + i\Gamma)$ (where $\omega_0$ and $\Gamma$ are the resonance frequency and width), it readily follows that the reflectivity, Eq. (1), is maximum at resonance for all angles except for those at which the incident angle coincides with the dipole orientation $\varphi = \pm\theta$:

$$R(\omega = \omega_0, \theta = \pm\varphi) \sim 0, \qquad R(\omega = \omega_0, \theta \neq \pm\varphi) \sim 1. \tag{3}$$

Specifically, the latter result (namely, R=1 at resonance) follows from a detailed calculation involving the quasi-analytical calculation of the dressed polarizability.

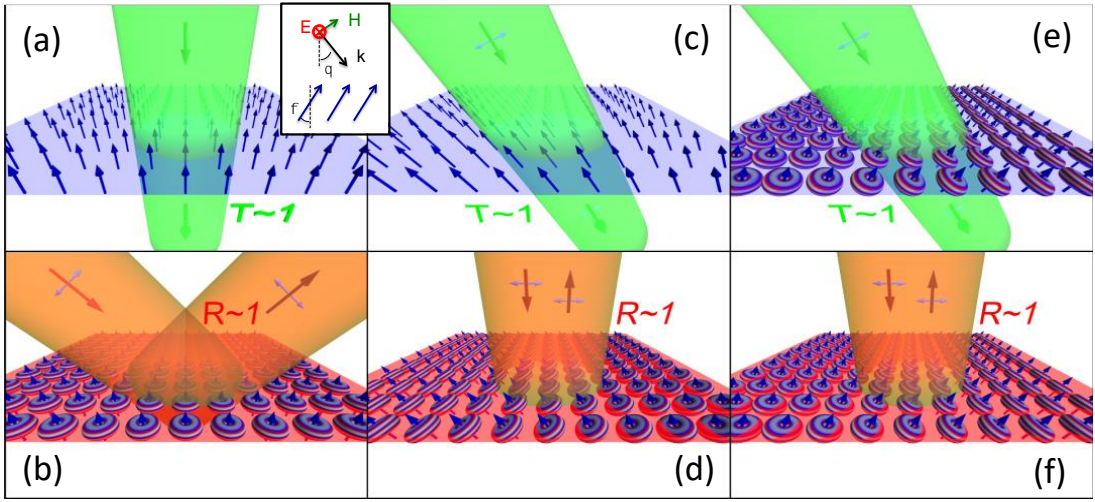

**Figure 1.** Brewster quasi-BIC concept schematic: Reflectance and transmittance through a metasurface of electric or magnetic dipolar scatterers at resonance; note that polarization vectors (blue arrows) are depicted (where needed) that refer to electric/magnetic fields depending on the electric/magnetic character of the dipoles. Dipolar emission patterns are included only when resonance excitation takes place. (**a,b**) Perpendicular dipole array showing a Brewster BIC at normal incidence (**a**, *T*~1), while exciting a resonant (leaky mode) at oblique incidence (**b**, *R*~1). (**c-f**) Oblique dipole array showing a quasi-Brewster BIC at oblique incidence with the same angle (plus or minus) such that: (**c**) it coincides with the direction of the dipole moments, so that coupling into dipole resonant modes is forbidden (*T*~1); (**e**) dipole resonances are indeed excited, but reflection is (Brewster-effect) forbidden (*T*~1 due to energy conservation, although dipole emission in transmission is not forbidden). (**d,f**) At any other incidence, e.g. at normal incidence, the resonant (leaky) mode is excited, leading to *R*~1.

This indicates that a Brewster-like quasi-BIC appears at $(\omega_0, \pm\varphi)$ due to the symmetry-protecting mechanisms: angular locking to a single scattering channel by the array periodicity, and forbidden scattering at a given angle determined by the single dipole resonance orientation. Actually, when $\varphi = 0$ (perpendicular dipole array, see Fig. 1a,b), this is strictly verified and a proper BIC with diverging Q factor emerges (*T*~1). However, when $\varphi \neq 0$ (oblique dipole array), the canonical BIC becomes a quasi-BIC with large, but finite, Q factors, since outgoing channels are allowed. These quasi-BICs nonetheless exhibit a rich phenomenology, making them inaccessible through plane wave illumination at given angles (see Fig. 1c-f), as follows. In turn, there is a slight difference depending on the sign of the angle of incidence, $\theta = \pm\varphi$, so that only scattering into reflection/transmission, respectively, is forbidden; reflection is identical due to reciprocity, but transmission reveals subtle differences. When the incident wavevector coincides with the resonant dipole orientation ($\theta = -\varphi$, Fig. 1c), coupling is forbidden and all the energy is directly transmitted through (*R*~0, *T*~1). Conversely, if the specular reflection direction coincides with the dipole orientation ($\theta = \varphi$, Fig. 1e), the incident wave couples into dipole resonances: however, dipole emission in reflection is forbidden (*R*~0), but not in transmission, which, in spite of that, leads to total transmission (*T*~1, Fig. 1e) thanks to incident wave cancelation. At any other angle of incidence ($\theta \neq \pm\varphi$), resonant excitation and emission is allowed and the wave is totally reflected (*R*~1, Fig. 1b,d,f). To summarize, a quasi-BIC with slightly smaller Q factors appears at oblique incidence that becomes inaccessible at given (Brewster) angles.

Note that we use the term "quasi-BIC" in a slightly different manner as was employed before[8,35]. In recent works, this term has been coined to indicate two phenomena that reduce the expectedly infinite Q-factor: the impact of finiteness of the system on a proper BIC, which makes spurious outgoing radiation plausible[8]; and high (but finite) Q-factor (not fully bound) eigenmodes emerging in asymmetric dimer metasurfaces[35], which only become true BICs with infinite Q-factors when dimers become symmetric[35,36]. The Brewster quasi-BICs that appear in the configuration proposed above resemble the latter quasi-BICs in the fact that their intrinsic Q-factors are finite, but differ from them since they are not accessible through plane wave excitation at a given angle due to a Brewster-like cancellation; this is in turn related to the very appealing property that they allow only for a single, specific outgoing channel. This feature has been so far exploited for BIC-induced lasing[7,13]: in both cases, quasi-BICs (not termed so therein, though) have been used instead of canonical BICs, exhibiting finite (but large) Q-factors allowing for an outgoing channel needed for lasing.

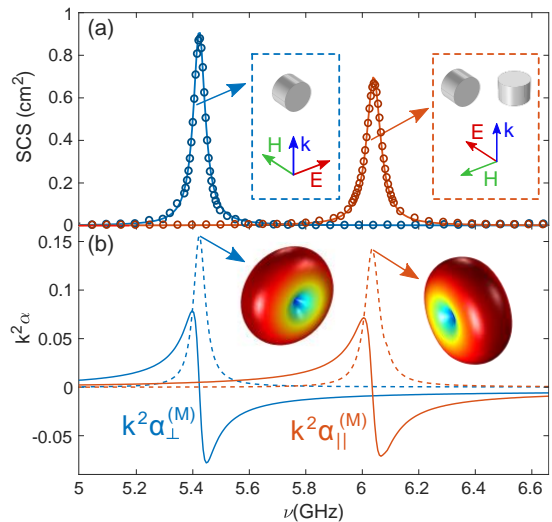

**Figure 2.** Experimental and numerical SCS of a single HRI disk reveals a non-degenerate MD resonance. (**a**) Scattering cross sections of a disk (r=3 mm and L=4 mm) with dielectric constant of $\varepsilon=78+i0.05$ for different incident wave polarization and disk orientations: SCUFF numerical results in $cm^2$ (solid curves) and experimental measurements in a.u. (symbols) are shown; the latter are obtained from experimental scattering spectra (see Supplemental Fig. S1) through angular integration in a dipole scattering plane (rather than in the entire space). Insets show the scattering geometry for all resonances. (**b**) Real and imaginary parts of the MD polarizibilities extracted from the numerical calculations that will be used below in the coupled electric/magnetic dipole formulations. Insets show the toroidal shape of the dipolar radiation pattern oriented in accordance with the scattering geometries shown in the insets in (a).

**Single magnetic-dipole resonance HRI disks**
In order to fabricate a suitable metasurface, we search for a meta-atom that exhibits as a lowest-order Mie-like resonance a non-degenerate dipole resonance pointing only at a fixed direction. To this end, we use nearly lossless HRI disks, i.e. a cylindrical dielectric resonator with dielectric constant of $\varepsilon=78+i0.05$ (in the GHz spectral regime of interest), radius r=3 mm, and length L=4 mm. This cylindrical resonator has the same volume as a sphere of radius $R_s$=3 mm. In the considered frequency range ($\nu \in$ [5-6.6] GHz, $\lambda \in$ [45.5-60] mm wavelength range), it corresponds to size parameters $q=2\pi R_s/\lambda=0.3-0.4$. Our numerical results for the scattering cross sections (Fig. 2) reveal that the lowest-order resonances are indeed magnetic-dipole (MD) resonances at 5.4 and 6.1 GHz ($q$=0.34 and 0.38), and do not overlap at all with the electric-dipole resonance, which actually lies at higher frequencies (~8 GHz, not shown in Fig. 2). These two MD resonances result from (see insets in Fig. 2): the (non-degenerate) one at 5.4 GHz with a MD pointing along the cylinder axis, and the (doubly-degenerate) resonance at 6.1 GHz whose MD can actually point along any direction within the plane perpendicular to the cylinder axis. Incidentally, the numerical results are also exploited to extract the corresponding magnetic polarizibilities that will be used below in the coupled-electric-magnetic dipole formulations. The experimental results (see Methods) for a single disk are shown in the Supplementary Figures S1 & S2, including the angular/spectral patterns and phases. It should be emphasized that they are in remarkable quantitative agreement with the numerical simulations, keeping in mind that the overall dimensions of the cylinder are of the order of a tenth of the incoming wavelength.

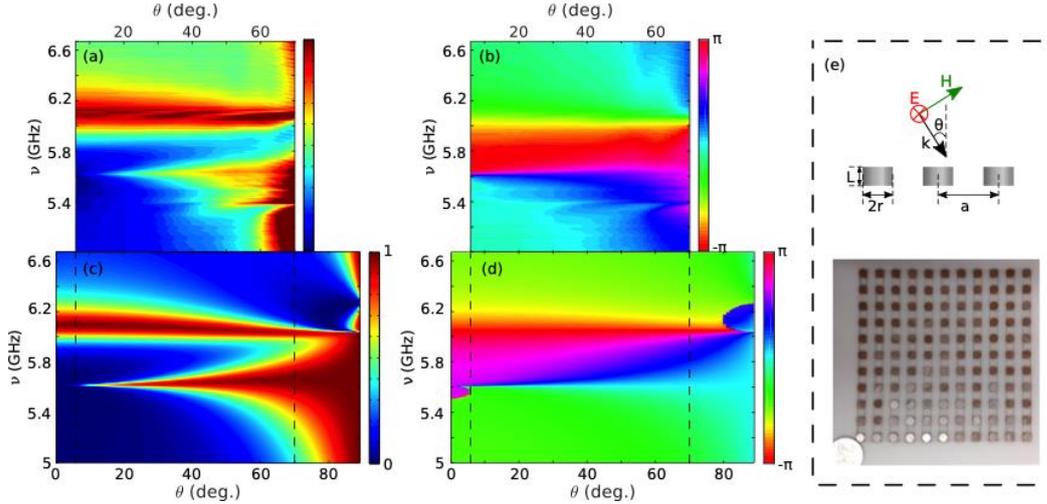

**Figure 3.** HRI disk metasurface reflectance showing Brewster BIC through experimental and theoretical spectra. Contour maps of the s-polarized reflectance R($\nu$,$\theta$) (**a, c**) intensity and (**b, d**) phase from a square array (lattice period=12 mm) of dielectric resonator disks (as in Fig. 2) as a function of angle of incidence $\theta$ and frequency (GHz). (**a,b**) Experimental measurements with AAC (see Methods) of the specular reflection (**a** intensity and **b** phase) for a finite array of 11x11 disks. (**c,d**) iCEMD theoretical calculations of the reflectance (**c** intensity and **d** phase) for an infinite array (see Methods). **e** Schematic depicting the geometry of the plane of incidence and picture of the sample.

**HRI Disk metasurface reflectance: Brewster quasi-BICs**

Next, we investigate the collective behavior of such MD-resonant disks in a metasurface. To this end, a finite square array of 11x11 disks (r=3 mm and L=4 mm) with lattice constant a=12 mm (as illustrated in the inset in Fig. 3) has been fabricated. Spectral and angular dependence of the measured reflectance intensity and phase are shown in Fig. 3a,b for TE (or S) polarization with an Apparent Angle Compensation (AAC), see Supplementary Fig. S6 & S7. The spectral and angular ranges ensure that there is no other diffraction order apart from the specular one (0th-order). Numerical simulation results have been also carried out (shown in Supplementary Figures S3-S4), calculated for an infinitely extended metasurface through SCUFF[47,48] (see Methods). Moreover, we show in Fig. 3c,d the theoretical calculations based on a coupled electric/magnetic dipole (iCEMD) theory[46] for an infinite metasurface (see Methods). Disks are indeed subwavelength, and can be replaced by one perpendicular MD and two (degenerate) parallel MDs. An excellent agreement with the experimental results is observed when the AAC is applied (despite the fact that calibration is lost in the process). This in turn indicates that the finite metasurface contains enough unit cells as to reproduce the reflectance of the infinitely extended system (except for the diverging theoretical Q-factor), so a convergence study is not strictly needed. A finite CEMD calculation (see Methods) for a finite (11x11 dipole) metasurface identical to that in the experiment is shown in the Supplementary Figure S5 to show that the results are directly comparable to those for finite arrays.

Apart from the large reflectance at grazing incidence through most of the spectral range, two spectral features weakly dependent on the angle of incident become evident: two total reflection bands at ~5.6 GHz and ~6.1 GHz. The latter one is broader at normal incidence, becoming slightly narrower at grazing angles. Conversely, the lowest-energy band is much broader at grazing incidence, becoming narrower for decreasing angle of incidence, and finally disappearing at normal incidence. Both bands stem from the single-disk MD resonances, slightly blue-shifted due to lattice coupling (as discussed below): perpendicular/out-of-plane (respectively, parallel/in-plane) for the one at 5.6 GHz (respectively, at 6.1 GHz), referred to as MD$_\perp$ (respectively, MD$_\parallel$). In addition, phase maps in Fig. 3b,d show an abrupt phase change along the resonant bands, with a more complex behavior around normal incidence for the MD$_\perp$ band, and around grazing incidence for the MD$_\parallel$ band.

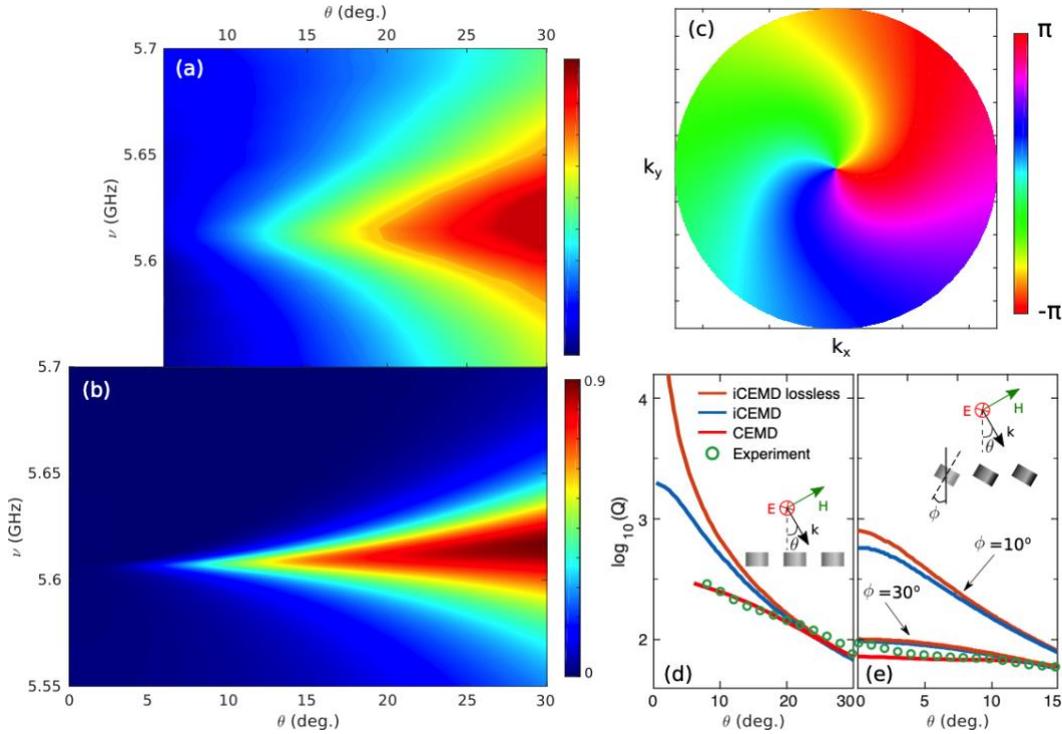

**Figure 4.** Symmetry-protected Brewster BIC characterization through experimental and theoretical spectra. (**a,b**) Contour maps of the S-polarized reflectance R($\nu,\theta$) intensity as in Fig. 3a,c from a square array (lattice period=12 mm) of dielectric resonator disks (as in Fig. 2), zooming in the BIC region around $\nu$=5.6 GHz and $\theta$=0º: (**a**) Experimental measurements for a finite 11x11 disk array; (**b**) iCEMD theoretical calculations for an infinite metasurface. (**c**) Phase map of the BIC vortex. (**d**) Q factors at resonance ($\nu$=5.6 GHz) obtained from the theoretical (reflectivity) and experimental (extinction cross section) spectral full width at half maximum in a and b, as a function of the angle of incidence $\theta$; the Q-factors from the finite CEMD calculation of the extinction spectra are included too. (**e**) Same as in d, but for arrays of tilted disks with tilt angle $\phi$=10º (only theoretical iCEMD) and $\phi$=30º (see Fig. 5).

Let us thus examine the MD$_\perp$ band around normal incidence, which exhibits clear features of a BIC: This is done in Fig. 4. First, we zoom in such region in Fig. 4a,b and explicitly show in Fig. 4d the experimental and theoretical Q-factors (Q=$\nu/\Delta\nu$): It is evident from them that the Q-factor increases exponentially when approaching normal incidence. Indeed, it should be emphasized that the experimentally measured Q-factor is extremely large, Q≥300, ideally tending to infinite as shown in the iCEMD calculations in the absence of losses within numerical accuracy (Q~2000 if losses are included). To give even more evidence, the topological nature of such MD⊥-BIC is explored by showing its associated vortex state in *k*-space[4,9]. Phase maps are shown in Fig. 4c calculated from the resulting numerical results for the reflectance electric field amplitudes as in ref.[4]: clearly, a vortex-like feature is reproduced with an associated topological charge of 1.

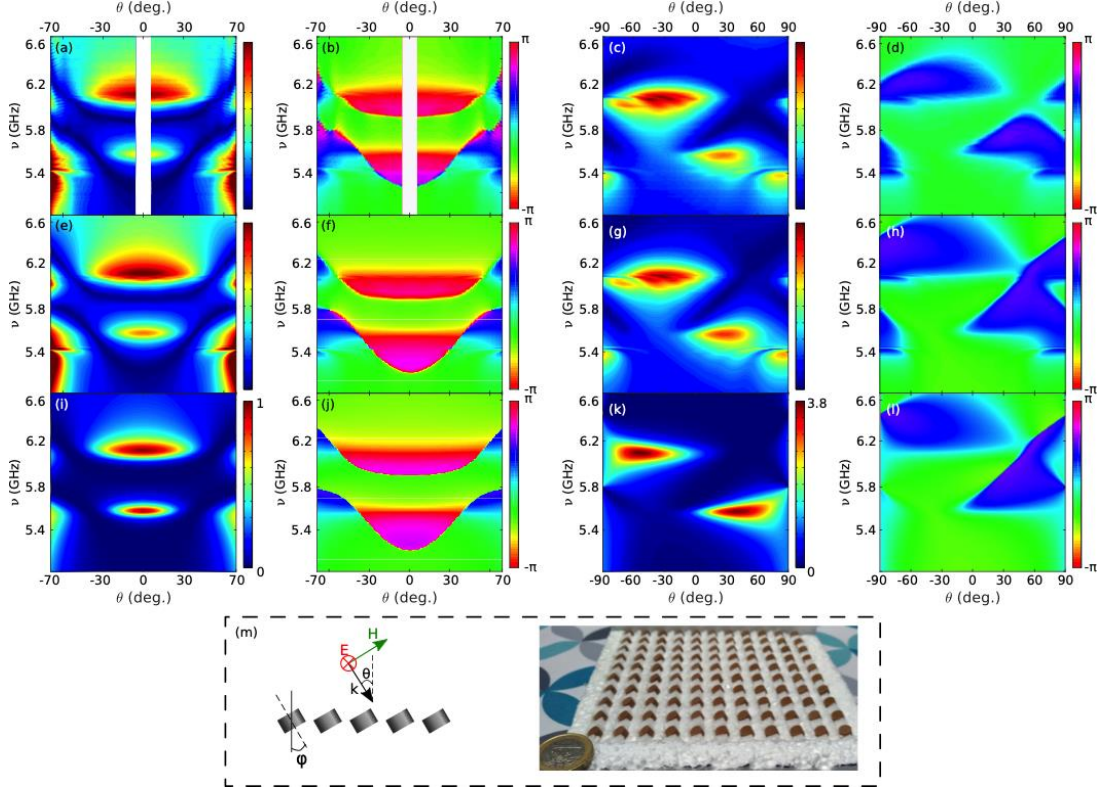

**Figure 5.** Angularly-tuned Brewster quasi-BICs in HRI tilted-disk metasurface: experimental and theoretical spectra. Contour maps of the s-polarized reflectance R(ν,θ) intensity and phase from a square array (lattice period=12 mm) of dielectric resonator disks (as in Fig. 2), tilted ϕ=30º with respect to the plane normal, as a function of angle of incidence θ and frequency (GHz). **(a-d)** Experimental measurements of intensity with AAC and phase for a finite 11x11 array of disks: **(a,b)** reflection (excluding the non-accessible backscattering region); **(c,d)** zero-order contribution into transmission, removing the incident field from the total transmitted field (see Methods). **(e-h)** same as in a-d, but for CEMD theoretical calculations (with AAC). **(i-l)** same as in a-d, but for the iCEMD theoretical calculations for an infinite array. **m** Schematic depicting the geometry of the plane of incidence and picture of the sample.

Therefore, a clear BIC associated to the MD⊥ band arises at normal incidence that is symmetry-protected through a Brewster-like effect: such perpendicular MD⊥ resonance cannot emit along the MD direction, which is however the only (specular) direction of reflection allowed by the metasurface lattice geometry. On the other hand, there are also some features that suggest that the MD∥ band approaches a "virtual" BIC at grazing incidence (see Fig. 3), supported by a similar argument (even though strictly speaking there should not be such BIC right at the light line, since it would become a guided mode). Next, let us explore how such symmetry-protected BICs evolve at non-zero/non-grazing angles of incidence when both MDs were rotated.

Accordingly, we have fabricated a metasurface with 11x11 disks, tilted $\phi=30°$ with respect to the plane normal, as shown in the inset in Fig. 5; this has been done as commented in the Methods section. Experimental measurements are plotted in Fig. 5a-d. In this case, since the number of disks is even smaller than in the non-tilted disk metasurface due to sample fabrication limitations, comparisons with theoretical calculations for both finite (CEMD, with AAC as in the measurements) and infinite (iCEMD) metasurfaces are included in Fig. 5e-l. Measurements without AAC (for both non-tilted and tilted metasurfaces) are given in Supplementary Figs. S5-S6 for both polarizations, along with finite CEMD calculations for the same finite disk metasurfaces. Note that the quantitative agreement between measurements and finite CEMD calculations is remarkable in all cases. Moreover, except for slightly broader resonances and minor discrepancies near grazing angles where AAC actually enlarges noise (for a thorough discussion on the impact of AAC, see Supplementary Figs. S7-S8), they resemble in turn the infinite iCEMD calculations. This confirms the fact that our finite samples suffice to account for the underlying physics and resulting phenomenology expected for an infinite metasurface.

Both measurements and theoretical calculations of reflectance intensity (see Figs. 5a,e,i) reveal that the $MD_\perp$ (respectively, $MD_\parallel$) resonance bands vanish approximately at $\theta=\pm30°$ (respectively, $\theta=\pm60°$), which are the symmetry-protected (lattice-imposed) specular reflection angles at which the corresponding single-disk MDs cannot emit in reflection. Phase maps in Figs. 5b,f,j show a sign change at fixed frequencies, namely, those of both MD resonances. Note that Brewster bands[37,46,49] (total transmission) appear that cross the MD resonance bands at the corresponding quasi-BICs; phase jumps clearly identify the Brewster band boundaries.

Interestingly, it should be noted that the Brewster-induced suppression of reflectance does not strictly hold for transmittance, as commented above with regard to Figs. 1c,d. To clarify this issue, we have included in Fig. 5c,d,g,h,k,l the metasurface contribution into specular transmission, calculated/measured by removing the incident field amplitude from that of the total transmitted field. Brewster-induced suppression of reflectance at the $MD_\perp$ resonance ($\nu\sim5.6$ GHz) would occur strictly speaking for a symmetric angle of incidence, $\theta=30°$ (Fig. 1e); nonetheless, $MD_\perp$ dipole excitation is allowed, so that such dipoles may contribute into specular transmission, as evidenced in Figs. 5c,g,k at $\theta=30°$ and $\nu\sim5.6$ GHz, with phase sign changes in Figs. 5d,h,l. Energy conservation is preserved then due to the fact that dipole emission in transmission partially cancels the incident wave directly transmitted (similarly to the optical theorem argument). By contrast, incident wave coupling into the $MD_\perp$ resonance is totally forbidden at $\theta=-30°$ (Fig. 1c); there is no resonant dipole emission whatsoever (neither in reflection nor in transmission) and total transmission occurs since the incident wave is transmitted unaltered (evidenced by negligible metasurface contribution to transmission at negative $\theta=-30°$ and $\nu\sim5.6$ GHz in Figs. 5c,g,k, with constant phases in Figs. 5d,h,l). Nonetheless, reciprocity enforces that both cases yield exactly the same reflectance, as revealed above through a symmetric reflectance for $\pm\theta$ in Figs. 5a,e,i. Likewise, the same argument holds for the $MD_\parallel$ resonance band at $\nu\sim6.1$ GHz for angles of incidence $\theta=-(\pm60°)$, with an obvious sign change in the angle of incidence due to the fact that disk tilting by $\phi=30°$ leads to a $MD_\parallel$ orientation change of $-60°$. For the sake of completeness, numerical simulations are also included in the Supplementary Fig. S4, in excellent agreement with our iCEMD theory.

Therefore, MD-induced quasi-BICs emerging in HRI disk metasurfaces can become transparent at different angles of incidence tuned by tilting the disk with respect to the metasurface plane. As discussed above, we term this scenario as Brewster quasi-BICs: basically, they stem from the symmetry-protected BIC at normal incidence for non-tilted disks, becoming quasi-BICs with finite Q-factors for tilted disks and/or at off-normal incidence. Interestingly, a Brewster-like mechanism imposes that such quasi-BICs are not accessible at θ=±ϕ. The resulting modes show non-diverging Q~$10^2$-$10^3$ factors (see Fig. 4e), that decrease as the tilt angle increases as compared to those obtained at normal incidence for the symmetry-protected BIC (see Fig. 4d). Despite being non-diverging, the possibility of exploiting such quasi-BICs to enhance strong resonant modes with a tunable single (or none) outgoing channel makes them especially attractive for variety of enhanced optical processes, as is the case of a recent demonstration of a (different) quasi-BIC-based directional lasing[13].

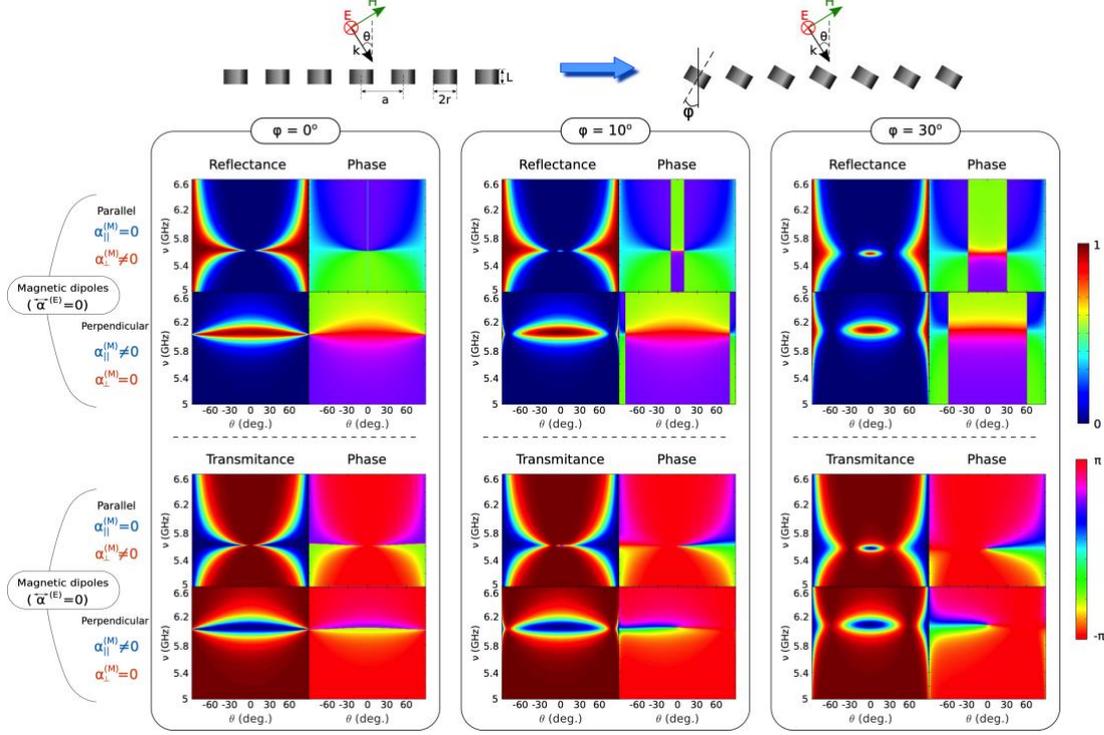

**Figure 6.** Brewster quasi-BIC theoretical model based on coupled electric & magnetic dipole array. Contour maps of the S-polarized reflectance (**first and second rows**) and transmittance (**third and fourth rows**) intensities/phases from an infinite square array (lattice period=12 mm) of dielectric resonator disks (as in Fig. 2), tilted ϕ=0°, 10°, 30° (**first to third groups of columns**, respectively) with respect to the plane normal, as a function of angle of incidence θ and frequency (GHz), theoretically calculated through iCEMD, including separately the contributions from each MD polarizability: out-of-plane (**first and third rows**) and in-plane (**second and fourth rows**).

To shed light on the physics underlying such quasi-BICs, we now make use of our iCEMD theory for infinite arrays mentioned above[46]. Contour maps of the reflectance (intensity and phase) as a function of frequency and angle of incidence are shown in Fig. 6, including both MD contributions separately, for 3 angles of incidence (columns): ϕ=0°, 10°, 30°. Contour maps of the transmittance (intensity and phase, third & fourth rows) are also included; absorption losses (not shown) are nearly negligible except for a very narrow band close to the MD resonances near BICs.

Upon inspecting the MD contributions at ϕ=0°, 10°, 30°, it becomes evident that symmetry-protected BICs appear for the MD⊥ (respectively, MD∥) resonances at θ=0° (respectively, at θ=90°), becoming Brewster quasi-BICs at θ=10°, 30° for the MD⊥ resonances (respectively, at θ=80°, 60° for the MD∥ resonances). A broadening near the Brewster BIC condition is observed for both MD⊥ and MD∥ resonances as the tilt angle increases, becoming as discussed above quasi-BICs. Phase maps (note that they differ from vortex maps) actually reveal Brewster BICs through a cross pattern with two abrupt phase jumps (sign change) delimited by: a horizontal line at resonance (already mentioned) and a vertical line at fixed Brewster angle. Our results also account for the lattice-induced shift of the single-disk MD resonances to 5.6 GHz and 6.1 GHz.

When all contributions are accounted for, the tails of each MD have a slight impact on the positions of the other MD-BIC, slightly shifting them to a different angle, as was indeed observed in the experimental and theoretical results shown above (Figs. 3 and 5). In addition, it is evident from the phase maps shown also in Figs. 3 and 5 that the vertical Brewster-induced phase jump bends as a result of the interaction between both MD resonances, as expected, whereas the horizontal jumps related to resonance positions remain unaltered.

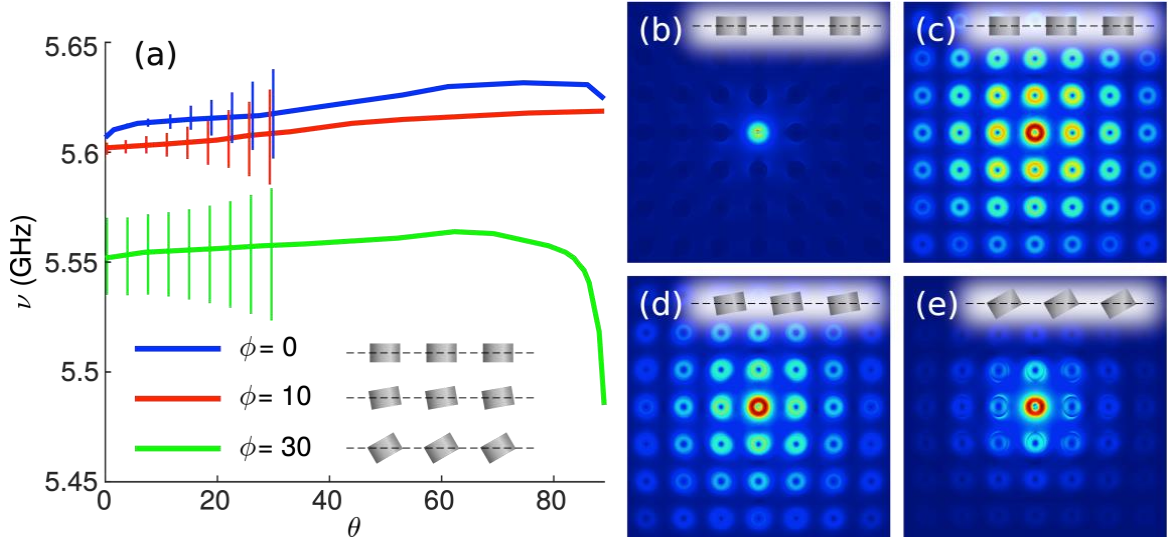

**Figure 7.** Brewster quasi-BIC theoretical analysis of eigenmodes. **(a)** iCEMD calculations of eigenmode dispersion relations (solid curves) and half-widths (error bars, up to 30°). **(b-d)** Numerically calculated (COMSOL) contour maps of electric near-field intensities on a horizontal plane crossing the disks through the center ($z=0$) for a 11x11 finite square array (lattice period=12 mm) of dielectric resonator disks, upon excitation with a MD located at the center of the central disk as follows (only shown the central 7x7 disk area): **(b,c)** vertical MD for non-tilted disks, **(b)** off-resonance ($\nu=5$ GHz) and **(c)** at resonance ($\nu=5.605$ GHz); **(d,e)** tilted MD at resonance ($\nu=5.605$ GHz) for equally tilted disks by an angle **(d)** 10° ($\nu=5.605$ GHz) and € 30° ($\nu=5.555$ GHz).

Finally, as yet another evidence of the existence of the proposed Brewster quasi-BICs, we make use of our iCEMD to calculate the eigenmodes with associated half-widths, solving for solutions of the homogeneous problem without incident plane wave. This is done in Fig. 4a for non-tilted and tilted disk metasurfaces. In both cases, eigenmode dispersion relation entirely coincides with the reflectance bands shown in Fig. 7a. For the non-tilted disk case, a canonical BIC is observed with diverging Q factor; however, as expected, a large (but finite) Q-factor is retrieved for the tilted disk metasurface. To further confirm their nature, numerical simulations have been carried out to reveal local electromagnetic fields for finite (though large) metasurfaces using on- and off-resonance excitation, see Fig. 7b-e. A vertical MD is placed inside the central disk, plotting the electric field intensity distribution on a horizontal plane parallel and very close to the metasurface. Off-resonance (Fig. 7b), the MD field is rapidly emitted, so that the near-field only concentrates in the excited disk. On-resonance, the MD field excites the true BIC (non-tilted disk, Fig. 7c) and the quasi-BICs (tilted disks, Fig. 7d-e), extending all of them over a large number of neighboring disks, as expected. Notably, both exhibit a similar decay away from the center, revealing that the quasi-BIC finite Q factor is large enough so that the finiteness of the samples has a larger impact on leakage; still, its Q-factor is larger the closer is the tilt angle to the vertical direction, as shown also in Fig. 4d,e.

## Conclusions

We have introduced the concept of Brewster quasi-BICs by combining the angular constraint to a single specular channel imposed by metasurface periodicity with the resonant behavior of single dipole, non-degenerate meta-atoms that radiate at all angles except for that of their induced dipole-moment orientation (Brewster effect). Such Brewster-like quasi-BICs have been experimentally shown to emerge from a symmetry-protected MD-induced BIC in HRI non-tilted disk metasurfaces. As the MD tilt-angle is increased, they become quasi-BICs, however being inaccessible at different angles of (plane-wave) incidence, which can be selectively tuned by tilting the disk axes with respect to the metasurface plane, in full agreement with coupled electric/magnetic dipole theories. In this manner, the angular and spectral position of the Brewster quasi-BIC is entirely controlled by the disk MD resonance orientation (tilted angle) and wavelength. Huge Q-factors are evidenced at normal incidence for perpendicular dipoles, while slightly smaller ones are obtained at oblique angles with opposite signs (reciprocal scenarios), termed quasi-BICs, due to the relaxation of the Brewster-condition either in reflectance or transmittance. This kind of BICs could be exploited in e.g. low-threshold mirrorless lasers[7,13], topological insulators[4,13,50,51], and other optoelectronic and sensing devices[52–56]. Bear in mind that, although the refractive indices of available semiconductor materials in the optical domain are not large enough as to spectrally separate dipolar resonances for nanospheres o nanostructures with aspect ratio close to 1, this spectral separability can nonetheless be accomplished by playing with shape and aspect ratio of meta-atoms[20]. Interestingly, the MD character of the proposed BICs may give rise in turn to intriguing new physics[57,58]. Moreover, the same universal mechanisms underlying such photonic Brewster quasi-BICs can be explored in any kind of waves.

## Methods

**Sample fabrication.** The disk of which were made different arrays is sold as a dielectric resonator. Its permittivity is given to be about 80, its radius 3 mm and its height 4 mm. The different arrays were made of ensembles of the previously mentioned disks positioned and spaced thanks to expanded polystyrene holders and spacers that have been properly machined.

**Measurement setup.** The experiments were made in the anechoic chamber of the CCRM used for those purposes by researchers of Institut Fresnel. This equipment has been already described in various papers. In the present study we measured the complex scattered field of a single disk and the reflection and transmission of different arrays from 3.5 to 8.5 GHz.

The scattered field of the single disk was obtained using a complex subtraction of the incident field (without the target) from the total field (with the target). This scattered field is then post-processed using a classical time gating and our drift compensation procedure. A calibration is also made using a reference target, here a metallic sphere, which is referenced to Mie computations. Thus, the resulting scattered field is determined fully quantitatively (considering an incoming wave of amplitude 1 V/m and null phase at the center of the target) and, furthermore, as it is determined in complex values, the experimental determination of the phases is fully achievable. The major difficulty of measuring one single disk is due to its small dimension: at 6 GHz its diameter is about $\lambda/8$ and thus its perturbation of the incoming wave is very small. This makes the complex subtraction of the measured incident field from the total field compulsory, but very sensitive to the dynamic and noise of the measurements, and to any drift phenomena (a few hundredths of radians must be compensated). The measurements of the scattered fields were made rotating the receiving antenna on a circle of about 4 m of diameter around the disk, this disk being placed on an expanded polystyrene mast, with permittivity very close to 1, that can be considered fully transparent to microwaves (more details can be found in Ref. 39 looking at the "horizontal configuration"). Its permittivity is very close to 1.

The reflection measurements of the arrays in the anechoic chamber were made using the vertical arch, moving simultaneously and symmetrically the source antenna and the receiver from the normal of the array (close to the "vertical" configuration in Ref. 39, see also Figs. S7-S8). The arrays were placed on the top of polystyrene mast and the same three steps measurement procedure (total, incident, reference fields) was used. The angular limitation of the travel of the antenna are, from one side, the length of the arch, and, from the other side, the physical dimensions of the antennas and of the wagons on which they are both fixed. The incident angle can thus be varied from 6º to 70°.

The transmission measurements of the arrays were made using two opposite antennas positioned in the azimuthal plane. This time, the antennas remain fixed while the array rotates on itself on the polystyrene mast. The same three steps measurement procedure (total, incident, reference fields) are performed in order to provide quantitative measurements.

For the reflection and transmission measurements, the same processing as for a single disk is performed on the measured fields, but the drift compensation cannot be applied because it is based on the knowledge of the full scattered field (or at least over a wide bistatic angular measurement) Fortunately, this compensation is not really needed here as the measured signals magnitudes are higher than those for a single disk, therefore less sensible to noise issues. The other specificity in those measurements processing is that we are comparing them to the computations obtained assuming either finite or infinite arrays. When the measurements are compared with finite arrays CEMD simulations, the comparison is straightforward and quantitative. But when comparing the measurements to computations of infinite arrays (through iCEMD) we had to propose an Apparent Angle Compensation (AAC). In fact, experimentally the apparent angle of the array seen from the source varies when the emitting antenna (or the array) rotates (see Supplementary Figs. S7-S8), which would not be the case with an infinitely large array. Such comparisons can therefore only be qualitative and are plotted in arbitrary units.

**Theoretical modeling: iCEMD & CEMD.** A coupled electric and magnetic dipole formulation has been developed for a 2D (planar) infinite array of in-plane (horizontal) and out-of-plane (vertical) electric and magnetic dipoles with arbitrary electric/magnetic polarizabilities (iCEMD). General quasi-analytical expressions are explicitly derived for reflection and transmission, similar to those reported for a 1D array[46]. Calculations for a finite array (see Supplementary Figs. S5-S6) were carried out through a classical coupled-dipole theory (CEMD) for a finite number of electric/magnetic dipoles. The corresponding electric and magnetic polarizabilities for our HRI mm-disks were extracted in all cases from the numerical scattering cross sections as shown in Fig. 2.

**Numerical simulations.** Numerical simulations through SCUFF[47,48] (open-source software package for analysis of electromagnetic scattering problems using the method of moments) have been carried out, in particular for: scattering cross sections and far-field angular patterns of individual disks; spectral and angular dependence of the reflectance from and transmittance through an infinite square array of HRI disks. In all cases, the (only) needed mesh file for the surface of the disk was obtained from that used for the COMSOL finite cylinder with a fine mesh.

## References

1. Hsu, C. W., Zhen, B., Stone, A. D., Joannopoulos, J. D. & Soljačić, M. Bound states in the continuum. *Nat. Rev. Mater.* **1**, 16048 (2016).
2. Hsu, C. W. *et al.* Observation of trapped light within the radiation continuum. *Nature* **499**, 188–191 (2013).
3. Bulgakov, E. N. & Sadreev, A. F. Bloch bound states in the radiation continuum in a periodic array of dielectric rods. *Phys. Rev. A* **90**, 53801 (2014).
4. Zhen, B., Hsu, C. W., Lu, L., Stone, A. D. & Soljačić, M. Topological Nature of Optical Bound States in the Continuum. *Phys. Rev. Lett.* **113**, 257401 (2014).

5. Wang, Q. *et al.* Optically reconfigurable metasurfaces and photonic devices based on phase change materials. *Nat. Photonics* **10**, 60–65 (2015).
6. Yoon, J. W., Song, S. H. & Magnusson, R. Critical field enhancement of asymptotic optical bound states in the continuum. *Sci. Rep.* **5**, 1–8 (2015).
7. Kodigala, A. *et al.* Lasing action from photonic bound states in continuum. *Nature* **541**, 196–199 (2017).
8. Taghizadeh, A. & Chung, I.-S. Quasi bound states in the continuum with few unit cells of photonic crystal slab. *Appl. Phys. Lett.* **111**, 031114 (2017).
9. Doeleman, H. M., Monticone, F., den Hollander, W., Alù, A. & Koenderink, A. F. Experimental observation of a polarization vortex at an optical bound state in the continuum. *Nat. Photonics* **12**, 397–401 (2018).
10. Timofeev, I. V., Maksimov, D. N. & Sadreev, A. F. Optical defect mode with tunable Q factor in a one-dimensional anisotropic photonic crystal. *Phys. Rev. B* **97**, 1–7 (2018).
11. Marinica, D. C., Borisov, A. G. & Shabanov, S. V. Bound States in the Continuum in Photonics. *Phys. Rev. Lett.* **100**, 183902 (2008).
12. Bulgakov, E. N. & Maksimov, D. N. Topological Bound States in the Continuum in Arrays of Dielectric Spheres. *Phys. Rev. Lett.* **118**, 267401 (2017).
13. Ha, S. T. *et al.* Directional lasing in resonant semiconductor nanoantenna arrays. *Nat. Nanotechnol.* **13**, 1042–1047 (2018).
14. Hu, Z. & Lu, Y. Y. Resonances and bound states in the continuum on periodic arrays of slightly noncircular cylinders. *J. Phys. B At. Mol. Opt. Phys.* **51**, (2018).
15. Zhang, W., Charous, A., Nagai, M., Mittleman, D. M. & Mendis, R. Extraordinary optical reflection resonances and bound states in the continuum from a periodic array of thin metal plates. *Opt. Express* **26**, 13195 (2018).
16. Abujetas, D. R., van Hoof, N., ter Huurne, S., Gómez Rivas, J. & Sánchez-Gil, J. A. Spectral and temporal evidence of robust photonic bound states in the continuum on terahertz metasurfaces. *Optica* **6**, 996 (2019).
17. Sadrieva, Z., Frizyuk, K., Petrov, M., Kivshar, Y. & Bogdanov, A. Multipolar origin of bound states in the continuum. *Phys. Rev. B* **100**, 115303 (2019).
18. Cong, L. & Singh, R. Symmetry-Protected Dual Bound States in the Continuum in Metamaterials. *Adv. Opt. Mater.* 1900383 (2019). doi:10.1002/adom.201900383
19. Kupriianov, A. S. *et al.* Metasurface Engineering through Bound States in the Continuum. *Phys. Rev. Appl.* **12**, 014024 (2019).
20. Rybin, M. V *et al.* High-Q Supercavity Modes in Subwavelength Dielectric Resonators. *Phys. Rev. Lett.* **119**, 243901 (2017).
21. Bogdanov, A. A. *et al.* Bound states in the continuum and Fano resonances in the strong mode coupling regime. *Adv. Photonics* **1**, 1 (2019).
22. Mukherjee, S., Gomis-Bresco, J., Pujol-Closa, P., Artigas, D. & Torner, L. Topological properties of bound states in the continuum in geometries with broken anisotropy symmetry. *Phys. Rev. A* **98**, 1–9 (2018).
23. Azzam, S. I., Shalaev, V. M., Boltasseva, A. & Kildishev, A. V. Formation of Bound States in the Continuum in Hybrid Plasmonic-Photonic Systems. *Phys. Rev. Lett.* **121**, 253901 (2018).
24. Rivera, N. *et al.* Controlling Directionality and Dimensionality of Radiation by Perturbing Separable Bound States in the Continuum. *Sci. Rep.* **6**, 1–7 (2016).
25. Meinzer, N., Barnes, W. L. & Hooper, I. R. Plasmonic meta-atoms and metasurfaces. *Nat. Photonics* **8**, 889–898 (2014).
26. Arbabi, A., Horie, Y., Bagheri, M. & Faraon, A. Dielectric metasurfaces for complete control of phase and polarization with subwavelength spatial resolution and high transmission. *Nat. Nanotechnol.* **10**, 937–943 (2015).
27. Kivshar, Y. All-dielectric meta-optics and non-linear nanophotonics. *Natl. Sci. Rev.* **5**, 144–158 (2018).
28. Holloway, C. L., Love, D. C., Kuester, E. F., Gordon, J. A. & Hill, D. A. Use of Generalized Sheet Transition Conditions to Model Guided Waves on Metasurfaces/Metafilms. *IEEE Trans. Antennas Propag.* **60**, 5173–5186 (2012).
29. Jain, A., Tassin, P., Koschny, T. & Soukoulis, C. M. Large Quality Factor in Sheet

30. Li, L. *et al.* All-dielectric metamaterial frequency selective surfaces based on high-permittivity ceramic resonators. *Appl. Phys. Lett.* **106**, 212904 (2015).
31. Li, L. *et al.* Achieving all-dielectric metamaterial band-pass frequency selective surface via high-permittivity ceramics. *Appl. Phys. Lett.* **108**, 122902 (2016).
32. Asadchy, V., Albooyeh, M. & Tretyakov, S. Optical metamirror: all-dielectric frequency-selective mirror with fully controllable reflection phase. *J. Opt. Soc. Am. B* **33**, A16 (2016).
33. Glybovski, S. B., Tretyakov, S. A., Belov, P. A., Kivshar, Y. S. & Simovski, C. R. Metasurfaces: From microwaves to visible. *Phys. Rep.* **634**, 1–72 (2016).
34. Cuesta, F. S., Faniayeu, I. A., Asadchy, V. S. & Tretyakov, S. A. Planar Broadband Huygens' Metasurfaces for Wave Manipulations. *IEEE Trans. Antennas Propag.* **66**, 7117–7127 (2018).
35. Koshelev, K., Lepeshov, S., Liu, M., Bogdanov, A. & Kivshar, Y. Asymmetric Metasurfaces with High-Q Resonances Governed by Bound States in the Continuum. *Phys. Rev. Lett.* **121**, 193903 (2018).
36. Abujetas, D. R., van Hoof, N., ter Huurne, S., Rivas, J. G. & Sánchez-Gil, J. A. From Fano resonances to bound states in the continuum in dipole arrays at THz frequencies. *arxiv:1901.03122* (2019).
37. Paniagua-Domínguez, R. *et al.* Generalized Brewster effect in dielectric metasurfaces. *Nat. Commun.* **7**, 10362 (2016).
38. García-Etxarri, A. *et al.* Strong magnetic response of submicron Silicon particles in the infrared. *Opt. Express* **19**, 4815 (2011).
39. Geffrin, J. M. *et al.* Magnetic and electric coherence in forward- and back-scattered electromagnetic waves by a single dielectric subwavelength sphere. *Nat. Commun.* **3**, 1171 (2012).
40. Evlyukhin, A. B. *et al.* Demonstration of Magnetic Dipole Resonances of Dielectric Nanospheres in the Visible Region. *Nano Lett.* **12**, 3749–3755 (2012).
41. Kuznetsov, A. I., Miroshnichenko, A. E., Fu, Y. H., Zhang, J. & Luk'yanchuk, B. Magnetic light. *Sci. Rep.* **2**, 492 (2012).
42. Paniagua-Domínguez, R., Abujetas, D. R. & Sánchez-Gil, J. A. Ultra low-loss, isotropic optical negative-index metamaterial based on hybrid metal-semiconductor nanowires. *Sci. Rep.* **3**, 1507 (2013).
43. Kuznetsov, A. I., Miroshnichenko, A. E., Brongersma, M. L., Kivshar, Y. S. & Luk'yanchuk, B. Optically resonant dielectric nanostructures. *Science* **354**, aag2472 (2016).
44. Aoni, R. A. *et al.* High-Efficiency Visible Light Manipulation Using Dielectric Metasurfaces. *Sci. Rep.* **9**, 6510 (2019).
45. Vaillon, R. & Geffrin, J. M. Recent advances in microwave analog to light scattering experiments. *J. Quant. Spectrosc. Radiat. Transf.* **146**, 100–105 (2014).
46. Abujetas, D. R., Sánchez-gil, J. A. & Sáenz, J. J. Generalized Brewster effect in high-refractive-index nanorod-based metasurfaces. *Opt. Express* **26**, 31523 (2018).
47. Reid, M. T. H. & Johnson, S. G. Efficient Computation of Power, Force, and Torque in BEM Scattering Calculations. *arXiv:1307.2966*
48. Reid, M. T. H. http://homerreid.com/scuff-EM.
49. Liu, W. & Kivshar, Y. S. Generalized Kerker effects in nanophotonics and meta-optics. *Opt. Express* **26**, 13085 (2018).
50. Lu, L., Joannopoulos, J. D. & Soljačić, M. Topological states in photonic systems. *Nat. Phys.* **12**, 626–629 (2016).
51. Alù, A. Metamaterials: Topological order gets active. *Nat. Phys.* **13**, 1038–1039 (2017).
52. Carletti, L., Koshelev, K., De Angelis, C. & Kivshar, Y. Giant Nonlinear Response at the Nanoscale Driven by Bound States in the Continuum. *Phys. Rev. Lett.* **121**, 033903 (2018).
53. Han, S. *et al.* All-dielectric active photonics driven by bound states in the continuum. 1–14 (2018).
54. Abujetas, D. R., Sáenz, J. J. & Sánchez-Gil, J. A. Narrow Fano resonances in Si nanocylinder metasurfaces: Refractive index sensing. *J. Appl. Phys.* **125**, 183103 (2019).
55. Fan, K., Shadrivov, I. V. & Padilla, W. J. Dynamic bound states in the continuum. *Optica* **6**, 169 (2019).

56. Liu, M. & Choi, D.-Y. Extreme Huygens' Metasurfaces Based on Quasi-Bound States in the Continuum. *Nano Lett.* **18**, 8062–8069 (2018).
57. Taminiau, T. H., Karaveli, S., Van Hulst, N. F. & Zia, R. Quantifying the magnetic nature of light emission. *Nat. Commun.* **3**, 976–979 (2012).
58. Sanz-Paz, M. *et al.* Enhancing Magnetic Light Emission with All-Dielectric Optical Nanoantennas. *Nano Lett.* **18**, 3481–3487 (2018).

## Acknowledgements
J.A.S.G. J.J.S. and D.R.A. acknowledge partial financial support from the Spanish Ministerio de Ciencia, Innovación y Universidades (MCIU/AEI/FEDER, UE) through the grants LENSBEAM (FIS2015-69295-C3-2-P, FIS2015-69295-C3-3-P) and NANOTOPO (FIS2017-91413-EXP), and from the Ministerio de Educación, Cultura y Deporte through a PhD Fellowship (FPU15/03566). A.B. wants to express her gratitude to the University of Cantabria for her PhD grant. D.R.A. and J.A.S.G. acknowledge fruitful discussions with R. Paniagua-Domínguez pointing out the subtle features for oblique dipole arrays.

## Author contributions
D.R.A., J.J.S., and J.A.S.G. conceived the idea of the metasurface Brewster BIC. F.M., J.J.S., J.-M.G., and J.A.S.G proposed the experiment; J.-M.G. and F.M. designed the experimental setup; J.-M.G. performed all experimental measurements and post-processing. A.B., F.M. and D.R.A. carried out the calculations for an isolated disk; A.L. and A.B. carried out CEMD calculations for a finite array of disks; D.R.A. and J.A.S.G. conducted the iCEMD calculations and SCUFF numerical simulations for the infinite metasurface. J.A.S.G. and J.-M.G. co-wrote the manuscript. All authors discussed the results and commented on the manuscript.

## Additional information
Supplementary information is available for this paper at .

## Competing interests
The authors declare no competing interests.